\documentclass{article}
\usepackage{amssymb}
\usepackage{amsmath}
\usepackage{amsmath,amssymb,amsthm,amscd,graphicx}
\usepackage[latin1]{inputenc}

\def\Tr{\operatorname{Tr}}

\def\deri#1#2{\frac{d#1}{d#2}}

\begin{document}

\title{Optimal Control of One-Qubit Gates}

\author{
K.~M. Fonseca-Romero\thanks{e-mail: karenf@ciencias.unal.edu.co}, 
G. Useche-Laverde,\\
{\small Departamento de Física, Universidad Nacional de Colombia,
Bogotá, Colombia} \\ 
\and F. Torres-Ardila\\ 
{\small Departamento de Física, 
Universidad de los Andes, A.A. 4976, Bogotá, Colombia.}
}

 \maketitle

\begin{abstract}
We consider the problem of carrying an initial Bloch vector
to a final Bloch vector in a specified amount of time under
the action of three control fields (a vector control field).
We show that this control problem is solvable and therefore
it is possible to optimize the control. We choose the physically
motivated criteria of minimum energy spent in the control,
minimum magnitude of the rate of change of the control and 
a combination of both. We find exact analytical solutions.
\end{abstract}

\section{Introduction}
Recent advances in experimental physics allowing for manipulation and measurement of single quantum systems, have stimulated a flurry of investigations on the control of quantum systems, and more or less formal schemes have been advanced\cite{Rabitz,Viola,Jacobs,Vitali}. The conditions under which a given quantum system is completely controllable have been explored \cite{RamakrishnaPRA51,RamakrishnaPRA61,Schirmer} and some limits of quantum controllability  \cite{SchirmerLimits} have been found. Quantum control theory have several important applications including quantum state engineering \cite{Sang}, control of chemical reactions \cite{RabitzChemical,Umeda}, laser cooling of molecular degrees of freedom \cite{Tannor,SchirmerCooling}, quantum register initialization \cite{Long} and the fabrication of robust quantum memories \cite{Greentree}. 

A major application of the theory of quantum control is the subject of quantum computation.
The physical implementation of a quantum computer is 
a major challenge and many proposals\cite{proposals} 
including ion traps,
optical cavities, and quantum dots have been advanced.
Promissing practical implementations should be scalable,
facing the problem of heat dissipation which gets worse along
with the shrinking of the size of the proposed  physical 
system. In this work we address the problem of carrying an 
initial qubit (more precisely of an initial Bloch vector)
to an specified final qubit in a given amount of time
using the minimum amount of energy possible. This paper is
organized as follows: first we review the controllability of 
Bloch vector, and then we formulate and solve the problem of
optimal control under the criteria of minimum energy, minimum
energy derivative and a combination of both.

Two level systems adequately model many physical systems 
(spin 1/2, photon polarization, atoms in (quase)
monochromatic electromagnetic fields, etc), despite its simplicity.
A general 1-qubit gate can be represented as a  2-level
quantum system, and the most general Hamiltonian for such a
system can be written as $H=h_0 I + \vec{h}\bullet\vec{\sigma}$,
where $\vec\sigma$ are Pauli´s matrices, $h_0$ determines
the zero energy reference, and $\vec h$ is a classical vector.
Since we can use any $\vec h$ we want, and this is the field
we use to control our system, here on we call $\vec h$ our vector
control. We moreover assume $h_0=0$. 

\section{Bloch vector Controllability}
All of the information of the quantum state of a two 
level system is completely determined by its 
density matrix $\rho$, or equivalently by its Bloch vector 
$\vec s(t) = \frac{1}{2}\Tr (\rho \vec\sigma)$.
The dynamics of Bloch vector, given by
well known Bloch equation $\dot{\vec{s}}(t)=\vec{b}\times{\vec{s}}$,
where $\vec b=2\vec h/\hbar$, can be put in the more explicit
way $\dot{\vec{s}}=(b_x{\cal J}_x+b_y{\cal J}_y+b_z{\cal J}_z)\vec{s}$,
where the ${\cal J}$s are the rotation generators.
In this case we can formulate the problem of taking an initial state
$\rho_i$ to a final state $\rho_f$ in a specified
amount of time $T=t_f-t_i$. After rescaling, we take $t_i=0, t_f=1$. 
Remember that we assume that all three components of $\vec h$ 
(or of $\vec b$) are 
control fields. Thus, since we 
have all three rotation generators, the system is 
completely controllable in the sense that every rotation can be 
reached from the identity \cite{Schirmer,Albertini}. In other words, 
any  final vector
can be reached from any other initial vector (of the same length, 
or the same degree of mixture) in any finite time, provided there 
are no constraints on the size of the control fields. Moreover, the 
motion equation can be inverted \cite{Agapi} to give $ \vec{b} =
\vec{s}\times\dot{\vec{s}} s^{-2}- f(t) \vec{s}$, where $f(t)$ is an
arbitrary function, which shows that 
not only the control problem is solvable, but that it is solvable even 
when a path to realize such control is given. Moreover, even in this
event the solution is not unique.

\section{Optimal Control}
As mentioned before, the larger the fields used for control the greater
the amount of heat to be dissipated. Then, proved the complete 
controllability of the one qubit gate, it is meaningful asking which is 
the control vector field required to perform an arbitrary rotation
operation with the minimum amount of energy spent.
We address this problem an optimal ``classical'' control 
problem for the Bloch vector, that is we shall extremize the cost 
functional
\begin{equation}
S=\int_{t_0}^{t_f} dt 
\left\{
\frac{1}{2}\vec{b}\bullet\vec{b}
+\vec{\lambda}\bullet\left(\dot{\vec{s}}-\vec{b}\times\vec{s}\right)
\right\},
\end{equation}
\noindent where $\vec{b}=2\vec{h}/\hbar$, and $\lambda$ is a (vector) 
Lagrange multiplier, and the momentum corresponding to $\vec s$. 
Notice that the form of the cost functional above
is not arbitrary: were the vector control field a magnetic field acting
on a spin half system, or an electromagnetic field for a charged two
level system, the (electro)magnetic energy would have had the assumed
form. For the sake of simplicity we scale the variables to get
the following cost functional
\begin{equation}
S=\int_{0}^{1} dt 
\left\{
\frac{1}{2a}\vec{b}\bullet\vec{b}
+\vec{\lambda}\bullet\left(\dot{\vec{s}}-\vec{b}\times\vec{s}\right)
\right\},
\end{equation}
where the constant $a$ have been introduced to have $\vec s$ and
$ \vec \lambda$ in the same units.

Not all of the resulting Euler--Lagrange equations
\begin{equation} \label{elsistema}
\vec{b}=a \vec{s}\times{\vec{\lambda}},\quad
\dot{\vec{s}}=\vec{b}\times{\vec{s}},\quad
\dot{\vec{\lambda}}=\vec{b}\times{\vec{\lambda}},
\end{equation}
are true dynamical equations: the first equation is a constraint.
A few simple calculations show that the system of equations 
(\ref{elsistema}) possesses the following constants of motion
$s^2=\vec{s}\bullet\vec{s}$, $\lambda^2 = \vec{\lambda}\bullet\vec{\lambda}$
and $\nu = \vec{\lambda}\bullet\vec{s}=\lambda s \cos (\theta)$,
where $\theta$ is the angle between $\vec \lambda$ and $\vec s$. 
The equations of motion can be put in the equivalent form
\begin{equation}\label{sistemareducido}
\deri{}{[a t]}\begin{pmatrix}{\vec s}\\ {\vec \lambda}\end{pmatrix}=
\begin{pmatrix} -{\vec \lambda} \cdot {\vec s} & s^2
\\ -\lambda^2 & {\vec \lambda} \cdot {\vec s} \end{pmatrix}
\begin{pmatrix}{\vec s}\\ {\vec \lambda}\end{pmatrix}=
\begin{pmatrix} -\cos (\theta) & 1\\ -1 & \cos (\theta) \end{pmatrix}
\begin{pmatrix}{\vec s}\\ {\vec \lambda}\end{pmatrix},
\end{equation}
whose solution reads
\begin{equation}
\begin{pmatrix}{\vec s}\\ {\vec \lambda}\end{pmatrix}
=\frac{1}{\sin(\theta)}
\left(\begin{array}{cc} 
\sin(\theta-a t\sin\theta) & \sin(a t \sin\theta)\\
 -\sin(a t\sin\theta) & \sin(\theta+a t \sin\theta)
\end{array}\right)
\begin{pmatrix}{\vec s}(0)\\ {\vec \lambda(0)}\end{pmatrix}.
\end{equation}
\noindent We have assumed that $\vec s$ and $\vec \lambda$ are unitary
vectors. Notice that we have obtained an orthogonal transformation.
In fact this could have been anticipated by showing that $\vec{b}$ is 
constant both in norm and direction. 
Let us see that the solution is valid for any value of $\theta$, so we have
freedom to choose $\theta$ to our best convenience. For example, we can choose
$\vec \lambda(0)= \vec s (1)$. It follows $\cos \theta = 
\vec s (0) \cdot \vec s(1)$, and $\sin(a \sin(\theta))=\sin(\theta)$.
This choice leads to the solution
\begin{equation}\label{answer1}
\vec{s}(t) = \frac{\sin(b(1-t))}{\sin(b)}\vec{s}_i+
 \frac{\sin(bt)}{\sin(b)}\vec{s}_f,
\end{equation}
\noindent where 
\begin{equation}\label{answer2}
\vec b (t) = \vec b (0) =
\frac{{\rm ArcSin}\left(\sin (\theta)\right)}{\sin (\theta)}
{\vec{s}_i\times\vec{s}_f}= (\theta + 2\pi n)\vec{s}_\perp=
b_1(t;n)\vec{s}_\perp.
\end{equation}
\noindent Equations (\ref{answer1}) and (\ref{answer2}) constitute 
the solution to the proposed problem, and agree with the result of 
reference \cite{Butkovskiy}. Let us see that in the scaled variables 
the magnitude
of the field is essentially equal to the value of the angle between the 
initial and the final Bloch vectors. The solution corresponds to a rotation
around $(-)\vec{s}_\perp$ at constant speed, from the initial Bloch vector to
the final one. We see then that the solution is not 
unique: due to the 
multivaluedness of the function ArcSin, there is an infinite number of 
solutions. Each solution corresponds to a local minimum, the global minimum
being the solution whose magnitude is equal to the angle between the initial
and the final Bloch vector. The other solutions correspond to an integer 
number of turns followed by this angle. Finally, the other solutions 
correspond to reaching the final Bloch vector from the initial one but 
rotating the other way. Choices different to $\vec \lambda(0)= \vec s (1)$
are also meaningful, but should lie on the plane which contains the
initial and final Bloch vectors.
Let us see that the solution found breaks down
when the initial and final vector are antipodal. In this case there are
an infinite number of solutions all of them spending the same amount of
energy. In this case the choice of an initial Lagrange multiplier vector
which is not (anti)parallel to the initial Bloch vector
leads to a particular solution.

Now, we remark that the minimum energy control corresponds to minimum
path on the Bloch sphere.
The length $l$ transversed by the
tip of the Bloch vector is given by
\begin{equation}\label{geodesic}
l=\int_0^1 \left|  \deri{\vec s}{t} \right |dt =
\int_0^1 \left| {\vec  b} \times {\vec s}\right | dt =
s\int_0^1 b(t) |\sin(\theta(t))| dt \leq s\int_0^1 b(t) dt.
\end{equation}
This means that, for a given magnitude of the control field, the 
transversed length is maximized when the control
vector field makes a right angle with the Bloch vector. In other
words, if we restrict to control fields that are perpendicular
to Bloch vector, and if Bloch vector transverses a longer path, 
it is necessary to have a larger
average control field. This remark
allows us to restrict ourselves to consider the vector control to point
out in the direction of $\vec s_i\times \vec s_f$. Making the Ansatz
\begin{equation}\label{Eq:Ansatz4b}
\vec b(t) = b(t) \vec{s}_\perp,
\end{equation}
after some algebra we obtain
\begin{equation}\label{Eq:Solution4s}
\vec s(t) = 
\frac{\sin(\theta-\int_0^t b(t')dt')}{\sin (\theta)} \vec s_i+
\frac{\sin(\int_0^t b(t')dt')}{\sin (\theta)}\vec s_f,
\end{equation}
were $b$ should satisfy the equality $\int_0^1 b(t) dt =
\theta$. If we write $b(t)$ as $\theta + \delta(t)$ we see that the average 
value of $\delta(t)$ over the unitary interval is zero and that
$\int_0^1 b^2(t) dt = \theta^2 + \int_0^1 \delta^2(t) dt$.
Thus, the minimum is attained when the control vector field is
constant from the initial till the final time. 
It is possible to give an alternative argument which shows that 
the solution of minimum fluence is the same as the shortest geodesic.
Inverting the Bloch equation we obtain the control field
$\vec b = {\vec s}\times\dot{\vec s}+f {\vec s}$, and the energy spent
 in the control $\int_0^1  b^2(t) dt = \int_0^1  \left(
{(\dot{\vec s})^2-({\vec s}\cdot\dot{\vec s})^2+f^2}\right) dt$.
Since the second subintegral term is identically zero, and the function
$f(t)$ should be zero for the extrema, we see that the fluence minimization
and geodesic minimization (see Eq. \ref{geodesic}) are almost the same,
and reach their extrema together. Had we chosen the squared length 
instead of the length, both expressions would have been identical.

Since the times to perform quantum computation are generally short in
low dimensional condensed matter systems, which are the most promising
candidates, one also should analize possible limitations set by the 
rate at which control fields can be set. In particular, it is worth 
paying attention that the solution (\ref{answer2}) is a discontinuous one,
zero before the initial time, constant between the initial and final 
times, and zero again from the final time  on. Had we used the square 
of the time derivative of the control field
instead of the square of the field itself, defining the cost functional

\begin{equation}
S=\int_{0}^{1} dt 
\left\{
\frac{1}{2\Omega^2}\deri{\vec{b}}{t}\bullet\deri{\vec{b}}{t}
+\vec{\lambda}\bullet\left(\dot{\vec{s}}-\vec{b}\times\vec{s}\right)
\right\},
\end{equation}
the solutions obtained above
would have been also solutions of the new problem. 
In this case a whole set of new
solutions arise, which are of constant magnitude but whose direction
changes with time. It is easy to construct such a kind of solutions.
Let $\{\vec b(t), \vec s(t)\}$ a solution of the new problem, but with
a time dependent $b^2(t)$, then $\{\tilde{\vec b}(t) = 
\vec b(t)+f(t)\vec s, \vec s\}$, is also
a solution, no matter how the function $f$ is chosen. In particular,
we can adjust $f$ to obtain $\tilde b^2$ a constant. For instance,
if we set $\vec s (t) = \cos (\phi(t)) {\vec s}_0 (t)+
\sin (\phi(t)) {\vec s}_\perp$,
where ${\vec s}_0$ is the solution for the problem of minimum fluence,
and $\cos(\phi(t))$ a function with value $1$ both at $t=0$ and at $t=1$, 
we have the 
control field $\vec{b}(t)=\theta\cos^2(\phi)\vec{s}_\perp-
\dot{\phi}\vec{s}_\tau-\theta\sin(\phi)\cos(\phi)\vec{s}_0$ where 
$\vec{s}_\tau=\vec{s}_\perp\times\vec{s}_0$ is a unitary vector needed
to define a time dependent right triade $\{\vec{s}_0,\vec{s}_\tau,
\vec{s}_\perp\}$. One can choose $f$ as 
$f=\pm \sqrt{B^2-\theta^2\cos^2(\phi)-\dot{\phi}^2}$, with $B^2$
the maximum value of $b^2$, so at instants where the maximum is attained,
$f$ vanishes. For the sake of definiteness we use $\phi(t)=\theta\mu
t(1-t)$ which yields $f^2=\theta^2(\sin^2(\phi(t))+\mu^2t(2-t))$, and
produces a new constant norm vector control $\tilde{\vec{b}}$ with
magnitude $\theta\sqrt{1+\mu^2}$. This solution, of course, spends more
energy than the found before to perform the same control.

Solutions with vanishing magnitude at the initial and final instants
 of time also exist. 
Observe that in this case the equation for $\vec b$ is
\begin{equation}\label{Eq4b}
\ddot{\vec b}= -\Omega^2 {\vec s}\times{\vec \lambda}
\end{equation}
which is a true dynamical equation. This allows for some extra 
flexibility: now we can add initial and final conditions on the 
value of the control field. From the point of view of the energy
injected to the system, the most physically sensible conditions
are those of vanishing control field both at the initial and the
final times. We observe that in this case the solution should
follow the shortest geodesic between the initial and the final
Bloch vectors. In fact, since the control field begins and
ends with a vanishing value, it should grow and decrease as slowly
as possible but fast enough as to reach the maximum value necessary
to have an average magnitude of at least $\theta$. If it had grown to 
a greater value it would have grown at a larger pace, so it would have
been not the minimum solution sought.

Differentiating the equation (\ref{Eq4b}) we obtain
$\frac{d^3\vec{b}}{dt^3} = \vec{b}\times\frac{d^2\vec{b}}{dt^2}$, 
which immediately tells us that the second derivative of the vector
control has constant norm. 
The first integral of this third order equation, 
$ \frac{d^2\vec{b}}{dt^2}=\vec{b}\times\frac{d\vec{b}}{dt}-\Omega^2
\vec{s}_i\times\vec{s}_f$,
where we have assumed $\lambda(0)=\vec{s}_f$, 
can be solved under the assumption that the control
vector is  a second degree polynomial in $t$, with the result
$\vec{b}(t)=\frac{\Omega^2|\sin(\theta)|}{2} t(1-t)
\vec{s}_\perp =
b_2(t)\vec{s}_\perp$ . Numerical solution of these equations, without
the ansatz made above also lead to the same solutions.
Before expressing $\vec b$ completely in terms of ${\vec s}_i$
and $\vec{s}_f$, we proceed to discuss the more general physical
criterion in which one is interested on energy saving but with
a limited rate of change of the vector control, through the
cost functional
\begin{equation}\label{GeneralS}
S=\int_{0}^{1} dt 
\left\{
\frac{1}{2a}\left(
\vec{b}\bullet\vec{b}
+\frac{1}{\omega^2}\deri{\vec{b}}{t}\cdot\deri{\vec{b}}{t}
\right)
+\vec{\lambda}\bullet\left(\dot{\vec{s}}-\vec{b}\times\vec{s}\right)
\right\}.
\end{equation}
The experience gained with the previous examples shows that the 
solution control fied should point (anti)parallel to
$\vec{s}_\perp$. Some algebra leads to the
solution
\[
\vec{b}(t)=a|\sin(\theta)|
\left(1-\frac{\cosh\omega(t-\frac{1}{2})}{\cosh(\frac{\omega}{2})}\right)
\vec{s}_\perp =
b_3(t)\vec{s}_\perp.
\]
We notice that all of the solutions so far found have the form of
equation (\ref{Eq:Ansatz4b}) and therefore have the solution 
(\ref{Eq:Solution4s}). We only have to take care of the final
value of $\vec s$. This leads to the following more explicit forms for 
$b(t)$
\begin{equation}
b_2(t;n)=6(\theta+2\pi n) t(1-t),\qquad
b_3(t;n)=\frac{\theta+2\pi n}{1-\frac{\tanh(\omega/2)}{\omega/2}}
\left(
1-\frac{\cosh(\omega(t-\frac{1}{2}))}{\cosh(\frac{\omega}{2})} 
\right).
\end{equation}
For the second case considered, the intuitive choice, $b=\pi\theta
\sin(\pi t)/2$ produces a value of the cost functional only 1.5\%
above that of the optimal solution. 
Finally, for control fields of the form of equation 
(\ref{Eq:Ansatz4b}) the cost functional can be written in 
purely geometric terms.
If we set $\dot{\phi}(t)=b(t)$, $\phi(0)=0$, then $S$ can be expressed as
\begin{equation}\label{Eq:GeometricS}
S=\frac{1}{2a}\int_0^{\bar{b}} b(\phi)
\left(
1+\left(\frac{1}{\omega}\deri{b}{\phi}\right)^2
\right) d\phi,
\end{equation}
where $\bar{b}$ is the average magnitude of the control field, and 
$\phi$ the accumulated angle (or the arc length) transversed by 
the Bloch vector. Equation 
(\ref{Eq:GeometricS}), just like equation (\ref{GeneralS}), 
contains the other two cases: the first in the
limit $1/\omega\rightarrow 0$, and the second in the limit $1/a
\rightarrow 0$ but with $a\omega^2=\Omega^2$ fixed. Of course,
$b_1(t;n) = \lim_{1/\omega\rightarrow 0} b_3(t;n)$ and
$b_2(t;n) = \lim_{1/a\rightarrow 0, a\omega^2 = \Omega^2} b_3(t;n)$.

We have formulated and solved in an analytic way, the problem of 
rotation of the Bloch vector (which characterizes completely the 
state of a two level system) from a prescribed initial vector to
a prescribed final vector, in a given amount of time, using an
optimal control scheme which minimizes the energy spent by the
control fields, or the magnitude of the rate of change of the
control  fields or a linear combination of both. We have found
 control fields  perpendicular to both
the initial and final Bloch vectors, and multiple local minima 
corresponding
to arrival from the initial to the final Bloch vector in one or other
senses or after one or more complete turns.

This work was partly funded by  DIB-UN
(División de Investigaciones, Sede Bogotá, Universidad Nacional).

\end{document}